\documentclass[aps,twocolumn,floats,preprintnumbers]{revtex4-2}
\usepackage{graphics,graphicx,epsfig,color,epstopdf} 
\usepackage{gensymb}            
\usepackage{amsmath,amssymb}    
\usepackage{mathrsfs}           
\usepackage{bibunits}           
\usepackage{siunitx}            
\usepackage{subcaption}         
\usepackage[export]{adjustbox}  
\usepackage{wrapfig}
\defaultbibliography{refs}

\DeclareSIUnit\molar{\mole\per\cubic\deci\metre}
\DeclareSIUnit\Molar{\textsc{m}}
\DeclareMathOperator{\sinc}{sinc}
\begin{document}

\newcommand{\mytilde}{\raise.17ex\hbox{$\scriptstyle\mathtt{\sim}$}}

\title{Rheological dynamics of active \textit{Myxococcus xanthus} populations during development}
\author{Matthew E. Black}
\author{Joshua W. Shaevitz}
\email{shaevitz@princeton.edu}
\affiliation{Joseph Henry Laboratories of Physics and the Lewis-Sigler Institute for Integrative Genomics, Princeton University, Princeton, NJ 08544, USA}

\begin{bibunit}
\begin{abstract}
  The bacterium \textit{Myxoccocus xanthus} produces multicellular protective droplets called fruiting bodies when starved. These structures form initially through the active dewetting of cells into surface-bound droplets, where substantial flows of the material are needed as the fruiting bodies grow and become round. These dynamics are followed by a primitive developmental process in which the fluid-like droplets of motile cells mature into mechanically-resilient mounds of non-motile spores that can resist significant mechanical perturbation from the external environment. To date, the mechanical properties of fruiting bodies and the changes in cellular behavior that lead to maturation have not been studied. We use atomic force microscopy to probe the rheology of droplets throughout their development and find that relaxation occurs on two time scales, $\sim$1~s and $\sim$100~s. We use a two-element Maxwell-Wiechert model to quantify the viscoelastic relaxation and find that at early developmental times, cellular motility is responsible for the flow of the material but that this flow ceases when cells stop moving and become nonmotile spores. Later in development there is a dramatic increase in the modulus of the droplet as cells sporulate and the fruiting body matures, resulting in a mostly elastic structure that can protect spores from harsh environmental insult.
\end{abstract}

\maketitle

\begin{figure}
  \includegraphics[width=0.5\textwidth]{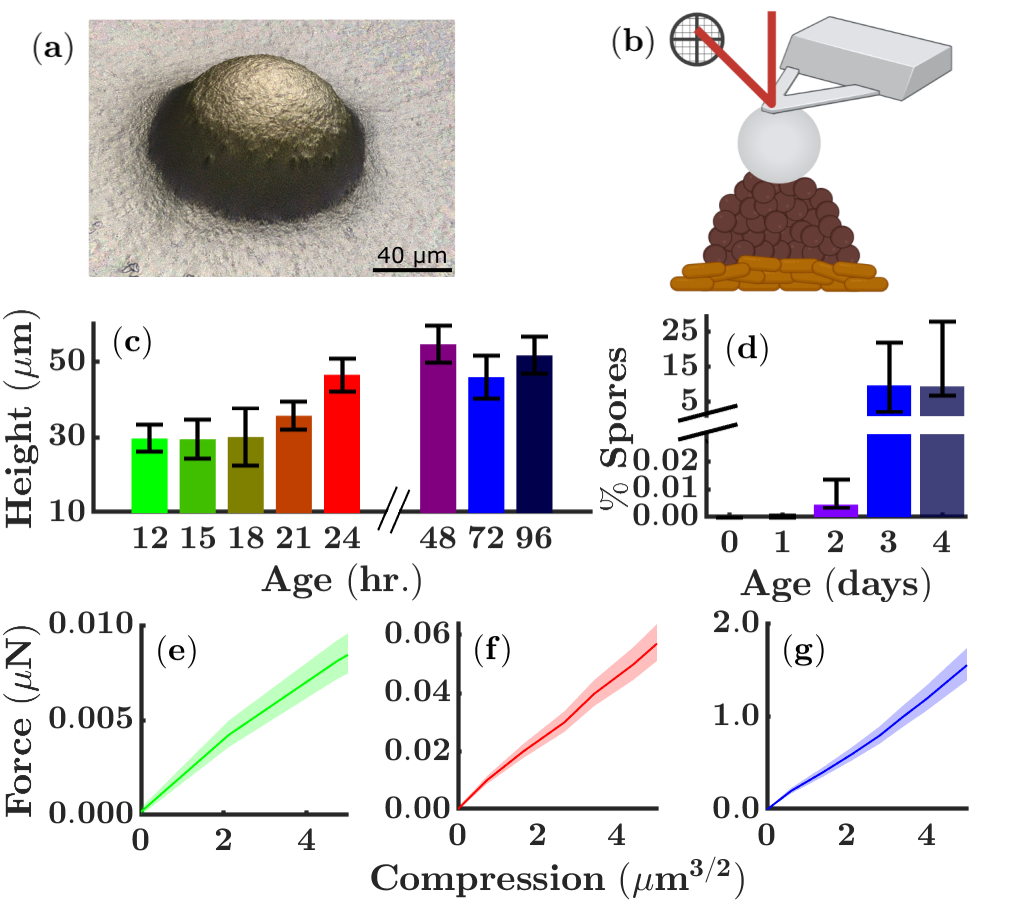}
  \caption{
    \textbf{(a)} Three-dimensional image of a typical fruiting body after 48 hours of starvation.
    \textbf{(b)} Experimental schematic. A bead approximately the same size as the droplet is attached to the end of an AFM cantilever to distribute the compressive forces. Created with BioRender.
    \textbf{(c)} Fruiting body height plateaus 24 hours post-starvation. 
    \textbf{(d)} Sporulation percentage plateaus 3 days post-starvation.
    \textbf{(e, f, g)} Force as a function of compression at difference times post-starvation (12~hr, green, $N=3$; 24~hr, red, $N=3$; 72~hr, blue, $N=4$).
  }
  \label{fig:main::introMyxo}
\end{figure}

Populations of the bacterium \textit{Myxococcus xanthus} form an active fluid that wets solid surfaces and allows cells to spread and predate on other species.
When food is scarce, this fluid dewets from the substrate to form droplet-shaped mounds inside of which a fraction of the population differentiates into chemically and mechanically resilient spores \cite{o1991development}.
The resultant macroscopic structure, a fruiting body, acts as a linchpin in the social lifestyle of \textit{M. xanthus} by maintaining the spore population as a coherent whole against physical and chemical stresses from the environment \cite{munoz2016myxobacteria,lee1994cloning}.
While considerable work has been done towards understanding the genetic underpinnings of fruiting body morphogenesis, a physical basis for how these structures form during the initial dewetting of the cell population, and the mechanical properties that allow them to fulfill their ultimate ecological function, remain unknown. 

To address this deficit, we used atomic force microscopy (AFM) to probe the rheology of \textit{M. xanthus} aggregates throughout their development from a dewetted droplet to a mature, spore-filled fruiting body (Fig. \ref{fig:main::introMyxo}a, b).
Fruiting bodies were grown on glass coverslips using the method of Kuner \& Kaiser \cite{kuner1982fruiting} (Supp. Mat. \ref{supp:methods::cellculture}).
Nascent droplets were first identifiable from the wetted cell population 12 hours after the onset of starvation in a nutrient-less buffer solution.

Fruiting body formation takes place in two phases; growth of the aggregate occurs first followed by sporulation of the cells.
The growth phase takes approximately 24 hours, as evidenced by a plateau in fruiting body height over time (Fig. \ref{fig:main::introMyxo}c). 
In contrast, sporulation didn't begin until day 2 and reached a stable maximum of $9.4 \pm 6.7\%$ of the cells after 3 full days in agreement with previous findings \cite{kroos1987expression, zusman2007chemosensory} (Fig. \ref{fig:main::introMyxo}d, \ref{fig:supp::spores7}).

We compressed fruiting bodies using an AFM on different time scales to measure the frequency-dependent mechanical response. 
We attached $80~\mu$m borosilicate glass beads to tipless AFM cantilevers so that the compressive forces imposed by the AFM were distributed across the droplet surface (Fig. \ref{fig:main::introMyxo}b). 
To minimize adhesion and other surface effects, beads were treated with a siliconizing reagent prior to their attachment to the AFM cantilevers.

Contact in our experiments is well-approximated as Hertzian, where the imposed force, $F$, scales linearly with sample compression, $\delta$, to the three-halves power (Fig. \ref{fig:main::introMyxo}e-g).
To account for the hemielliptical shape of each droplet, a small correction factor ($\beta \approx 0.97-0.99$) is introduced into the contact model \cite{johnson1987contact} (\S \ref{supp:methods::shape-characterization}). 
We further assume that the droplets remain incompressible throughout development.
Force and compression are thus related by,
\begin{equation}
  F = \frac{16\sqrt{R_c}}{9\beta}E\delta^{3/2},
  \label{eqn:main::hertz}
\end{equation}
where $E$ is the droplet's Young's Modulus and $R_c$ is the radius of the (uncorrected) spherical contact area between the cantilever-attached glass bead and the droplet under test. 

To confirm the relevance of this contact model, we measured the force-compression relationship for aggregates 12, 24, and 72 hours post-starvation in the linear-contact regime, $\delta \lessapprox R_c/10$ \cite{radmacher2007studying} (Fig. \ref{fig:main::introMyxo}e-g).
For each age, the force was proportional to $\delta^{3/2}$ as predicted by the Hertzian model (Eq. \ref{eqn:main::hertz}).
We find a dramatic increase in aggregate stiffness that occurs over the developmental process.
While 12-hour aggregates have apparent Young's moduli of $241 \pm 58~{\rm Pa}$ (Fig. \ref{fig:main::introMyxo}d), mature fruiting bodies are over two orders of magnitude stiffer, $E = 36.6 \pm 15~{\rm kPa}$ (Fig. \ref{fig:main::introMyxo}g).
Fully-dewetted, but not-yet-sporulated, aggregates at 24 hours displayed an intermediate stiffness (Fig. \ref{fig:main::introMyxo}f), suggesting that the growth process occurs alongside other processes that cause stiffening.

We performed two complementary sets of rheological experiments on aggregates throughout development to better understand this complex mechanical evolution. 
First, we adapted the methods of Huang and Mahaffy to measure the dynamic rheological response as a function of frequency \cite{huang2004measurements,mahaffy2004quantitative}.
\textit{M. xanthus} cells move at a speed of about one body length per minute and reverse their direction with a characteristic timescale of tens to hundreds of seconds \cite{liu2019self}.
We imposed oscillations with log-spaced frequencies between $\SI{1}{\milli\hertz}$ and $\SI{10}{\hertz}$ on developing droplets, allowing us to measure their complex modulus on timescales relevant to the motion of the constituent cells.

\begin{figure}
  \includegraphics[width=0.5\textwidth]{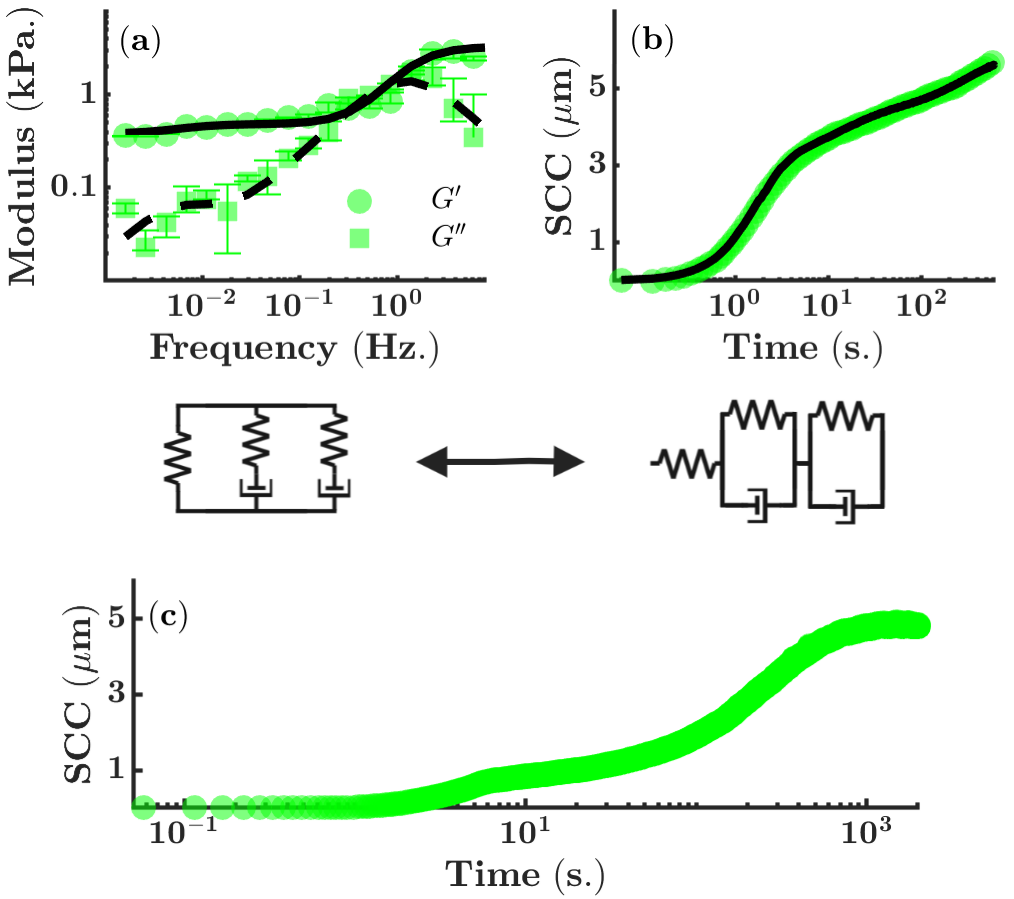}
  \caption{
    \textbf{(a)} The dynamic modulus as a function of frequency for a single 12-hr aggregate (green), and the resultant least-squares fit to the spring-dashpot model (black).  
    Error bars represent the smallest and largest moduli calculated from the fit coefficient confidence intervals.
    \textbf{(b)} 
    Compression (creep) time series for the same aggregate as in (a) under a constant compressive force of 60 nN (green), and the resulting fit to a viscoelastic spring-dashpot model (black). Compression is plotted as $SCC = \delta^{3/2}/\sqrt{R_c}$ (see text). 
    \textbf{(c)} Very long time-scale creep experiment demonstrating an equilibrium elastic modulus at long times.
  }
  \label{fig:main::exampleExpts}
\end{figure}

\begin{figure*}
  \centering
  \includegraphics[width=\textwidth]{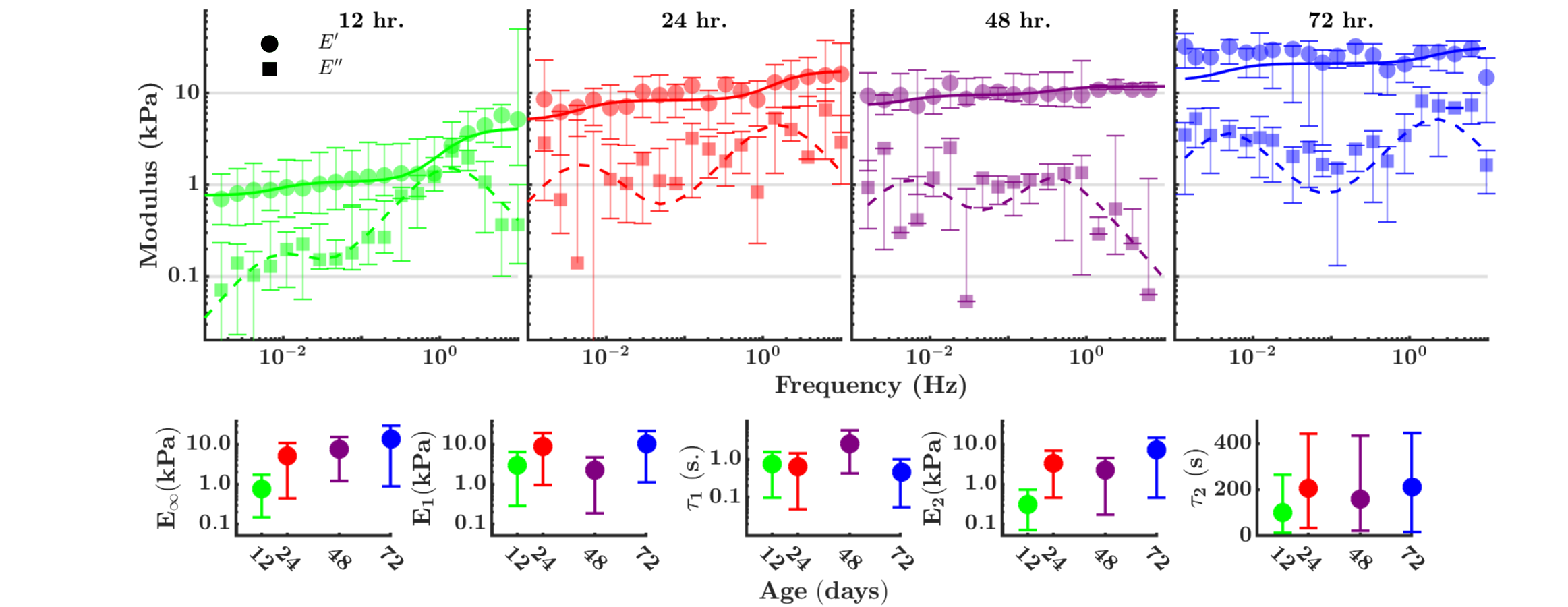}
  \caption{
    (Top) Dynamic modulus of fruiting bodies over the course of development. Points shown are median values for storage ($E'$) and loss ($E''$) moduli at each frequency, taken across all experiments. 
    $N = 4, 20, 6, 19$ for ages 12, 24, 48, and 72 hours post-starvation, respectively. 
    Lines are least-squares fits of the complex modulus over all frequencies to Eq. \ref{eqn:main::mwdynmod} (solid and dashed lines correspond with the storage and loss moduli, respectively).
    (Bottom) The change in each fit parameter is shown over the course of development.
    Error bars are bootstrapped 95\% confidence intervals.
    \label{fig:main::freqSweep} 
  }
\end{figure*}

\textit{M. xanthus} aggregates are viscoelastic (Figs. \ref{fig:main::exampleExpts}a, \ref{fig:main::freqSweep}).
Similar to our findings using the Young's modulus (Fig. \ref{fig:main::introMyxo}e-g), the storage moduli of developing aggregates increases by more than an order of magnitude over the course of development.
This change in modulus takes place without significant changes to the viscoelastic response, as evidenced by an apparent equilibrium modulus at long time scales and two, well-separated relaxation timescales at $\sim$1~s and $\sim$100~s for all ages (Fig. \ref{fig:main::freqSweep}).

This linear viscoelastic behavior can be described using a two-element Maxwell-Wiechert model with a complex modulus given by \cite{tschoegl2012phenomenological}
\begin{equation}
  E^{*}(\omega) = E_{\infty} + \frac{E_{1}\tau_{1}i\omega}{1 + \tau_{1}i\omega} + \frac{E_{2}\tau_{2}i\omega}{1 + \tau_{2}i\omega},
  \label{eqn:main::mwdynmod}
\end{equation}
where $E_{\infty}$ is the equilibrium modulus and the pairs ($E_1,\tau_1$) and $(E_2,\tau_2)$ are the moduli and relaxation times of the short- and long-time relaxation modes, respectively.
Least squares fits and values for model parameters are shown in Fig. \ref{fig:main::freqSweep} and Table \ref{tbl:supp::freqSweepParams}. We find that over the course of 72 hours, the equilibrium modulus, $E_\infty$, increases by more than an order of magnitude, along with large increases in $E_2$ and $\tau_2$.

While droplet viscoelasticity evolves throughout the developmental process, changes are especially large during the initial dewetting phase between 12 and 24 hours post-starvation.
The speed at which these changes take place, combined with the low frequencies where they occur, preclude us from making frequency-sweep measurements that avoid non-linear effects from sample evolution \cite{mours1994time}.
To probe aggregate viscoelaticity on short time scales, we performed time-domain creep experiments where a set force was imposed and held constant for a period of 10 minutes while the resulting compression was measured.
Guided by Eq. \ref{eqn:main::hertz}, we define a Size-Corrected Compression, $SCC \equiv \delta^{3/2}/\sqrt{R_c}$, and monitor its evolution over time (Fig. \ref{fig:main::exampleExpts}b,c).

Creep experiments resolved two relaxation timescales similar to the frequency-sweep experiments, but over a period short enough that each aggregate can be considered quasi-stable.
To extract mechanical model parameters from these experiments, the times series $SCC(t)$ for each experiment was fit to the Lee-Radok solution for viscoelastic contact with time-increasing contact area,
\begin{equation}
  SCC(t) = \frac{9\beta}{16}\int_{0}^{t}J(t-\tau)\frac{dF}{d\tau}d\tau,
  \label{eqn:main::leeradok}
\end{equation}
where $J(t)$ is the creep compliance function \cite{lee1960contact}.
To simplify the fitting routine, we converted the previously-described Maxwell-Wiechert model to its conjugate Kelvin-Voigt representation with creep compliance $J(t) = J_0(1 + q_1 (1 - e^{-t/\lambda_1}) + q_2 (1 - e^{-t/\lambda_2}))$ for use in in Eq. \ref{eqn:main::leeradok}.
The conjugate Kelvin-Voigt model exhibits a zero-frequency, ``glass'' compliance ($J_0$) and two retardation timescales $\lambda_1, \lambda2$ with accompanying compliances $J_{0}q_{1}$ and $J_{0}q_{2}$. 

Our frequency-sweep experiments predict that at very long time-scales the aggregates are solid-like.
However, this regime is not probed by the short-duration, 10-min creep experiments.
We confirmed a long-time, solid behavior by imposing a 60~nN force on 12 hour post-starvation droplets for 30 minutes and observed a second plateau in the $SCC$ at the expected long retardation timescale (Fig. \ref{fig:main::exampleExpts}c).

\begin{figure}
  \includegraphics[width=0.5\textwidth]{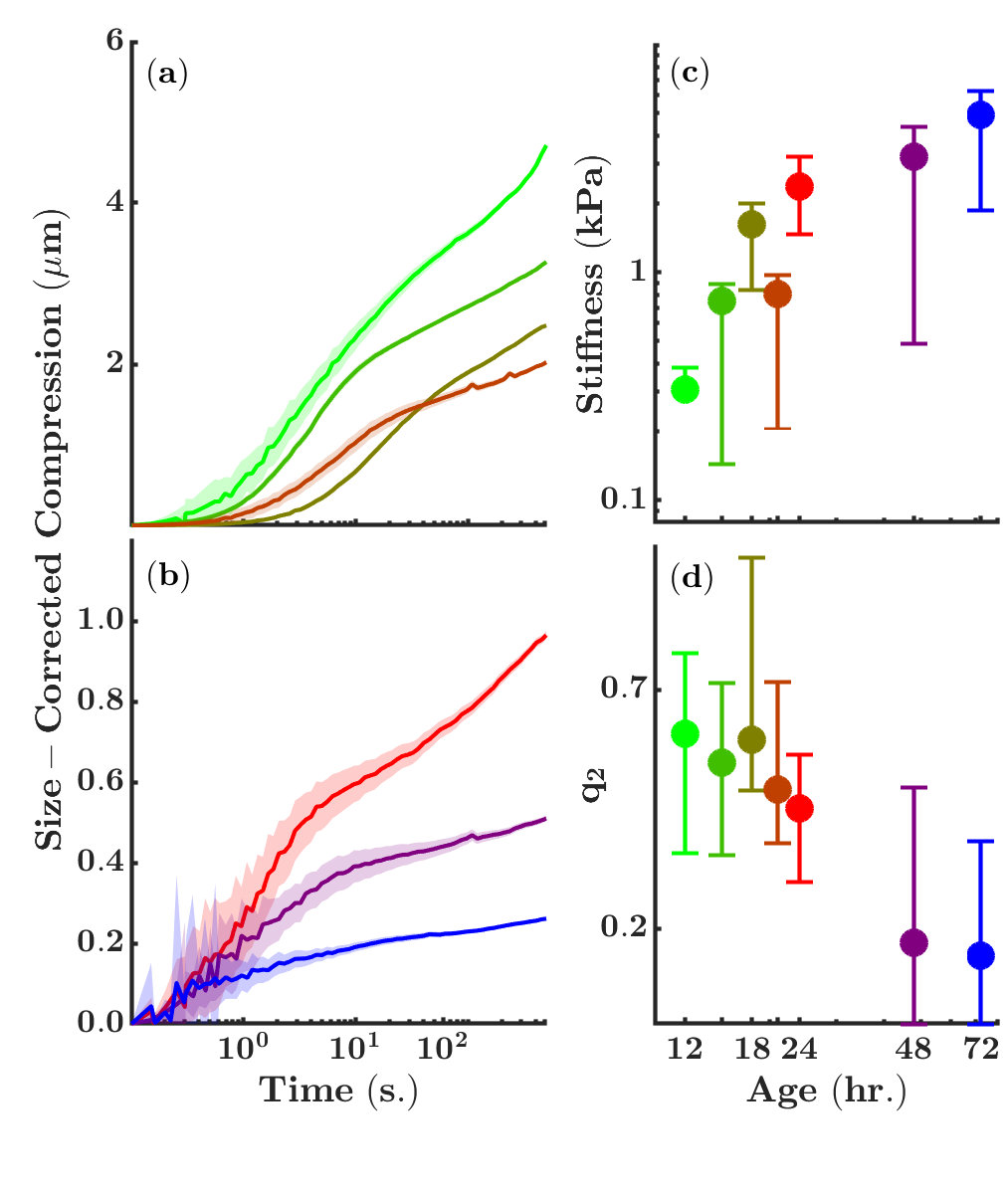}
  \caption{
    \textbf{(a,b)} Mean creep curves for early (a, 12, 15, 18, and 21 hours post-starvation, $N=13,4,7,7$, respectively) and late (b, 24, 48, 72 hours post-starvation, $N=12,10,13$, respectively) development. Colors are defined in panels c, d. Shaded regions are 95\% confidence intervals.
    \textbf{(c,d)} The stiffness, $J_0^{-1}$, and long timescale relative stiffness, $q_2$. Error bars are 95\% confidence intervals.
  }
  \label{fig:main::creepCurves}
\end{figure}

We performed creep experiments at 3 hour intervals during the initial dewetting period and found that changes in viscoelasticity over this period occur steadily. 
Tables \ref{tbl:supp::creepParamsEarly} and \ref{tbl:supp::creepParamsLate} describe the fitted model parameter values for fruiting bodies with ages 12, 15, 18, 21, 24, 48, and 72 hours.
Changes in both the zero-frequency stiffness, $J_0^{-1}$ (Fig. \ref{fig:main::creepCurves}c) and long-timescale compliance, $q_2$ (Fig. \ref{fig:main::creepCurves}d), occur at an approximately constant rate over this period.
Aggregates less than 24 hours post-starvation share a common response with flow attributable to the second retardation timescale, \textit{i.e.} occurring after approximately 10 seconds, producing an at-least doubling of the measured compression.
This  phenomenology matches the evolution of the complex modulus where both 12 and 24 hour post-starvation droplets show a marked reduction in both storage ($E'$) and loss ($E''$) moduli with inverse frequency.
In contrast, 48 and 72 hour old aggregates exhibit a near-constant storage modulus over the entire frequency range we probed.

Together, these data suggest that the mechanical evolution of developing fruiting bodies in the first 24 hours post-starvation and the subsequent two days are governed by different phenomena.
While the first day is dominated by a rapid stiffening of the droplet without suppression of flow on the minute-timescale, the mechanics of older droplets show less overall stiffening but an almost complete lack of creeping flow.
Strikingly, the cessation of flow after 24 hours occurs at the same time that droplets stop growing in size (Fig. \ref{fig:main::introMyxo}d).

\begin{figure}
  \includegraphics[width=0.5\textwidth]{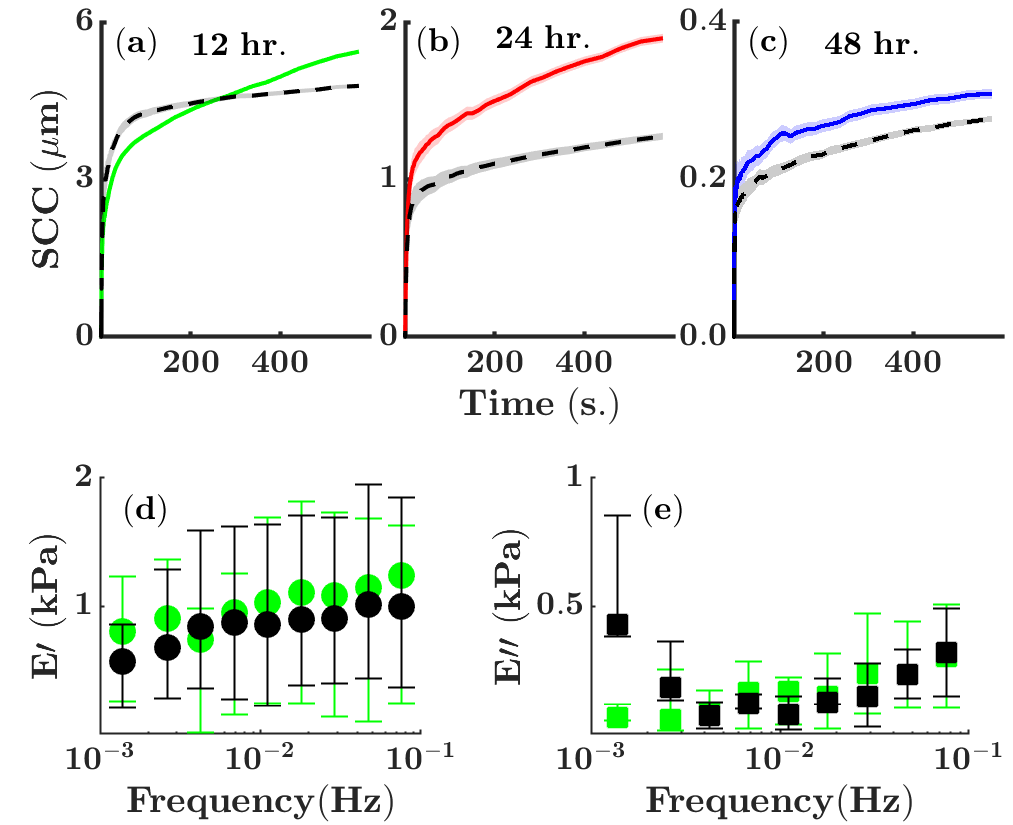}
  \caption{
    (a-c) Mean creep curves pre-treatment (colored/solid lines) and post-treatment (black/dashed) with the motility-halting drug CCCP. Shaded regions are 95\% confidence intervals. $N=4,7,5$ for (a,b,c), respectively. 
    Storage (d) and loss (e) moduli for 12 hour post-starvation droplets before (green) and after (black) CCCP exposure ($N=4$). 
  }
  \label{fig:main::cccpMotility}
\end{figure}

Droplet growth is driven by cellular motility, suggesting that cell motion generates viscous flows and fluidizes nascent droplets.
To test this hypothesis, we performed consecutive creep experiments on developing fruiting bodies before and after exposure to $\SI{20}{\micro\Molar}$ of the motility-halting drug carbonyl cyanide-m-chlorophenylhydrazone (CCCP).
CCCP is a proton ionophore that rapidly destroys the cellular proton motive force, causing motility to cease \cite{wartel2013versatile}.
Treatment with CCCP almost-completely suppressed the long-time flow of 12- and 24-hour old droplets (Fig. \ref{fig:main::cccpMotility}a,b).
CCCP-treated 12-hour old aggregates responded to mechanical perturbation similarly to 48-hour fruiting bodies in which the sporulation transition is underway and cell motility is naturally reduced.
Treatment with CCCP did not affect the retardation behavior of 48-hour post-starvation droplets (Fig. \ref{fig:main::cccpMotility}c).
While the exact mechanism for this flow suppression could not be quantified with this approach, complementary measurements of the dynamic modulus in 12-hour post-starvation droplets reveal an approximately five-fold increase in effective viscosity at low frequencies in the presense of CCCP (Fig. \ref{fig:main::cccpMotility}e). 

Taken together, our data provide the first mechanical characterization of \textit{M. xanthus} fruiting bodies during morphogenesis.
This process yields an aggregate that is dramatically stiffer than that of the initially-dewetted population.
Morphogenesis and stiffening take place in two steps, with growth of a viscous droplet occurring during the first day, and subsequent stiffening and solidification  during the following two days.
The  stiffening we observe takes place almost entirely as a result of changes in the long time, $\mathcal{O}(\SI{100}{\second})$, rheology of the aggregate.
The creeping flow at this long time scale is produced by cell motility as the aggregate forms.

There are several possible mechanisms that might lead to the large observed stiffening. 
First, spores are stiffer than vegetative cells and thus their presence might affect the stiffness of the fruiting body itself, although the relatively small percentage of cells that become spores likely indicates that this is not the sole mechanism. 
In addition to sporulation, \textit{M. xanthus} cells also excrete various polymeric substances, a mix of polysaccharides, proteins, and extracellular DNA termed the extracellular matrix (ECM) during development \cite{curtis2007proteins,hu2012dna}.
The ECM has been suggested to form a structural scaffold within the fruiting body (\cite{bonner2006fiba, lux2004detailed, arnold1988inhibition}). Future work will be needed to piece apart the molecular origins of the stiffening.

Here, we have shown how the initial fluid-like droplets formed by developing \textit{M. xanthus} populations transition to mature, solid-like fruiting bodies.
This work provides insight into the intimate link between mechanics and the biology of Myxobacteria, in addition to the physics of active solids.
While activity-induced viscosity reduction has been well-studied in dilute suspensions of motile bacteria \cite{martinez2020combined,saintillan2018rheology}, such an effect in dense, viscoelastic solids remains theoretically uncharacterized.

\vspace{\baselineskip} 
The authors thank Katherine Copenhagen, Endao Han, and Cassidy Yang for helpful discussions. This work was supported by the National Science Foundation through the Center for the Physics of Biological Function (PHY-1734030) and award PHY-1806501.

\putbib[Black_MyxoAFM2021_refs]
\end{bibunit}  

\widetext
\clearpage
\newpage

\begin{bibunit}
\begin{center}
  \textbf{\large Supplemental Material}
  
  \textbf{\large Rheological dynamics of active \textit{Myxococcus xanthus} populations during development}
\end{center}

\setcounter{equation}{0}
\setcounter{figure}{0}
\setcounter{table}{0}
\setcounter{section}{0}
\setcounter{page}{1}
\makeatletter
\renewcommand{\theequation}{S\arabic{equation}}
\renewcommand{\thefigure}{S\arabic{figure}}
\renewcommand{\thetable}{S\arabic{table}}
\renewcommand{\thesection}{S\arabic{section}}
\renewcommand{\bibnumfmt}[1]{[S#1]}
\renewcommand{\citenumfont}[1]{S#1}

\begin{table*}[h]
  \begin{tabular*}{0.74\textwidth}{ | l | c | c | c | c | }
    \hline
                                    & \textbf{12 hr.}     & \textbf{24 hr.}      & \textbf{48 hr.}      & \textbf{72 hr.} \\
    \hline
    $\bf E_{\infty}~\textbf{(kPa)}$ & 0.76 (0.38, 1.39)   & 2.89 (0.12, 7.34)    & 7.48 (6.86, 8.02)    & 13.6 (12.7, 16.0) \\
    $\bf E_{1}~\textbf{(kPa)}$      & 3.22 (1.79, 4.63)   & 7.67 (4.77, 13.92)   & 3.95 (1.62, 5.89)    & 10.4 (9.2, 11.4) \\
    $\bf \tau_{1}~\textbf{(s)}$     & 0.7 (0.55, 1.06)    & 0.55 (0.29, 0.95)    & 2.52 (2.10, 2.95)    & 0.45 (0.42, 0.52) \\
    $\bf E_{2}~\textbf{(kPa)}$      & 0.36 (0.10, 0.73)   & 5.06 (1.53, 12.67)   & 2.29 (2.12, 2.35)    & 7.29 (6.84, 7.63) \\
    $\bf \tau_{2}~\textbf{(s)}$     & 137.6 (54.1, 201.0) & 202.6 (166.6, 300.8) & 160.8 (138.4, 272.6) & 210.7 (195.9, 210.7) \\
    \hline
    $\textbf{N}$                    & 4                   & 20                   & 6                    & 19 \\
    \hline
  \end{tabular*}
  \caption{
    Median and (bootstrapped 95\% confidence intervals) for model parameters, extracted from least-squares fits of oscillatory experiments to Equation \ref{eqn:main::mwdynmod}, at each developmental age.
    \label{tbl:supp::freqSweepParams}
  }
\end{table*}

\begin{table*}[h]
  \centering
  \begin{tabular*}{0.89\textwidth}{ | l | c | c | c | c | }
    \hline
    & \textbf{12 hr.} & \textbf{15 hr.} & \textbf{18 hr.} & \textbf{21 hr.} \\
    \hline
    $\bf J_{0}~(\textbf{kPa}^{-1})$ & 3.25 (0.55, 3.81) & 1.34 (0.74, 1.64) & 0.61 (0.48, 0.81) & 1.24 (0.71, 1.48) \\
    $\bf q_{1}$ & \num{3.9e-13} (\num{2.5e-15}, 0.35) & 0.15 (\num{2.38e-20},\num{0.42}) & \num{1.2e-8} (\num{2.4e-13},\num{0.17}) & \num{7.4e-15} (\num{1.7e-15},\num{3.5e-12}) \\
    $\bf \lambda_{1}~\textbf{(s)}$ & 0.02 (0.01, 8.05) & 4.58 (0.01, 14.98) & 0.03 (0.01, 0.38) & 0.01 (0.01, 0.01) \\
    $\bf q_{2}$ &  0.61 (0.40, 0.76) & 0.55 (0.37, 0.71) & 0.59 (0.49, 0.98) & 0.49 (0.38, 0.72) \\
    $\bf \lambda_{2}~\textbf{(s)}$ & 240.44 (204.39, 295.07) & 143.85 (83.07, 257.05) & 330.69 (254.31, 723.9) & 302.69 (141.3, 1013) \\
    \hline
    $\textbf{N}$ & 13 & 4 & 7 & 7 \\
    \hline
  \end{tabular*}
  \caption{
    Median and (bootstrapped 95\% confidence intervals) for model parameters, extracted from least-squares fits of creep experiments to Equation \ref{eqn:main::leeradok} during the initial dewetting period ($<\SI{24}{\hour}$ post-starvation). 
    \label{tbl:supp::creepParamsEarly}
  }
\end{table*}

\begin{table*}[h]  
  \centering
  \begin{tabular*}{0.78\textwidth}{ | l | c | c | c | }
    \hline
    & \textbf{24 hr.} & \textbf{48 hr.} & \textbf{72 hr.} \\
    \hline
    $\bf J_{0}~(\textbf{kPa}^{-1})$ & 0.42 (0.27, 0.60) & 0.31 (0.18, 0.47) & 0.20 (0.09, 0.28) \\
    $\bf q_{1}$ & \num{1.5e-11} (\num{1.2e-12},\num{1.4e-10}) & \num{1.2e-13} (\num{4.3e-19},\num{1.0e-12}) & \num{4.8e-12} (\num{5.8e-14},\num{3.0e-10}) \\
    $\bf \lambda_{1}~\textbf{(s)}$ & 0.04 (0.01,0.07) & 0.04 (0.03, 0.06) & 0.03 (0.02, 0.05) \\
    $\bf q_{2}$ & 0.45 (0.33, 0.55) & 0.17 (0.03, 0.48) & 0.14 (\num{5.83e-11}, 0.49) \\
    $\bf \lambda_{2}~\textbf{(s)}$ & 114.69 (79.26, 201.95) & 209.85 (114.93, 632.55) & 293.39 (111.28, 605.86) \\
    \hline
    $\textbf{N}$ & 12 & 10 & 13  \\
    \hline
  \end{tabular*}
  \caption{
    Median and (bootstrapped 95\% confidence intervals) for model parameters, extracted from least-squares fits of creep experiments to Equation \ref{eqn:main::leeradok} for all timepoints $\ge\SI{24}{\hour}$ post-starvation. 
    \label{tbl:supp::creepParamsLate}
  }
\end{table*}

\begin{figure}[h]
  \includegraphics[width=0.3\textwidth]{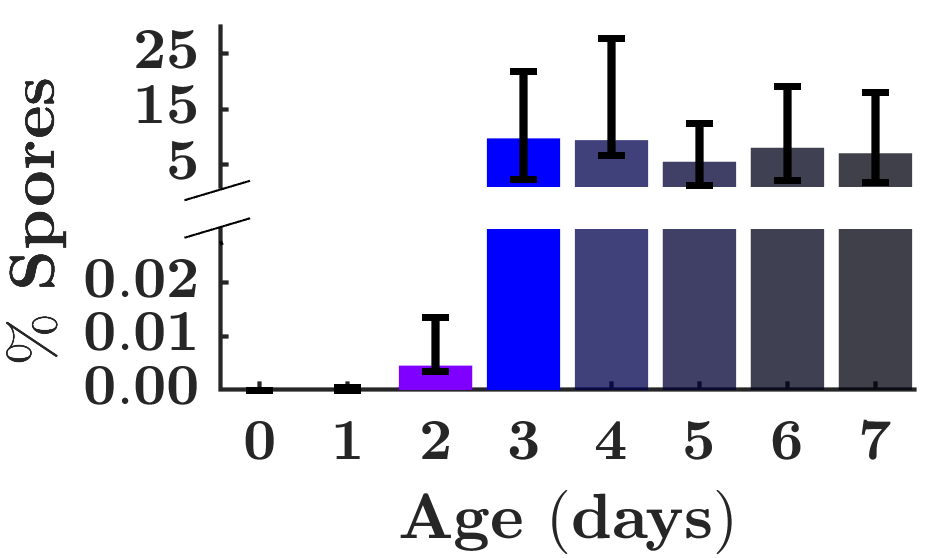}
  \caption{Percentage of the initial population that has sporulated at each day for one week after the onset of starvation (age=0). Measured as described in \ref{supp:methods::cellculture}.}
  \label{fig:supp::spores7}
\end{figure}

\begin{figure}[h]
  \includegraphics[width=0.5\textwidth]{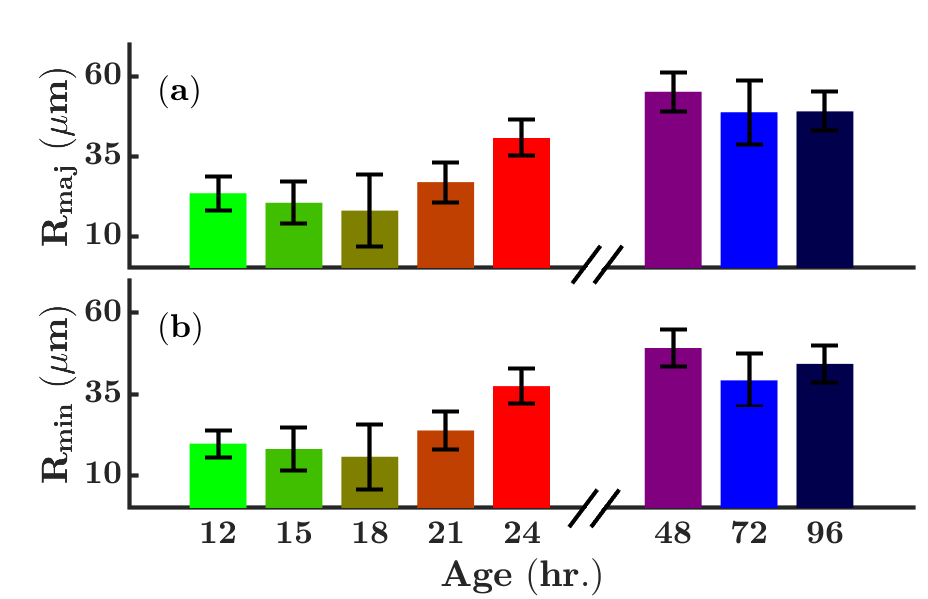}
  \caption{
    Measured major (a) and minor (b) radii of the droplet-base at each developmental age. 
    See \ref{supp:methods::shape-characterization} for details. 
  }
  \label{fig:supp::baseradii}
\end{figure}

\section{Materials \& Methods}
\label{supp:methods}

  \subsection{Cell Culture \& Fruiting Body Development}
  \label{supp:methods::cellculture}

  All experiments were performed with the \textit{M. xanthus} strain DK1622.
  Fruiting bodies were cultivated on glass coverslips by the method of Kuner and Kaiser \cite{kuner1982fruiting}. 
  Cells were grown shaking at 32\degree{}C in 1\% CTT medium (1\% Casitone, 10 mM Tris-HCl, 1 mM $\text{KHPO}_4$, 8 mM $\text{MgSO}_4$) to a density of approximately $1\times10^7$ cells/mL ($\text{OD}_{600} \approx 0.1-0.25$). 
  6 mL of the resultant culture was then centrifuged for 15 minutes at 1000 rcf on a tabletop Eppendorf 5424 centrifuge.
  The resulting pellet was resuspended in 4 mL of fresh CTT medium and transferred to a 60mm Petri dish (NEST Scientific, Cat No. 754001) to which \#1.5 coverslips (Globe Scientific) had been sealed to the bottom with Valap (1:1:1 mixture by weight of vaseline, lanolin, and paraffin wax).
  The dish was then placed in a static, 32\degree{}C incubator for 24 hours, allowing a thin biofilm to form across the bottom.
  Excess CTT was drained and the biofilm was washed by gentle rinsing with deionized water. Finally, the water was removed and the biofilm resubmerged in 4 mL of starvation media (10 mM MOPS buffer, 1 mM $\text{CaCl}_2$, buffered to pH 6.8 by addition of HCl as needed).
  We consider this time as the beginning of starvation (age = 0 hr.). 
  The dish was then placed back into a 32\degree{}C incubator until the relevant experimental time point was reached.

  Sporulation efficiency, or the percent of the total population that has sporulated, was measured using the sporulation assay described in \cite{whitworth2008myxobacteria}.
  Briefly, cells were manually scraped from the coverslip and suspended in $\SI{1}{\milli\liter}$ of sterile water.
  The suspension was then heated at 50\degree{}C while being sonicated for one hour in a Branson 1510 sonicator/water bath.
  Serial 10-fold dilutions were then prepared and plated in duplicate and allowed to growth for 5-7 days until single colonies were both visible and could be clearly demarcated by eye.
  Counts of these colonies were then compared to the initial cell concentration at time of starvation, measured by optical density of the initial culture, to compute the percent of the initial population that had sporulated at the time point under investigation. 

  \subsection{Atomic Force Microscopy}
  \label{supp:methods::afm-main}

  All experiments were performed on a custom-made atomic force microscope mounted on top of a Nikon TE-2000 inverted microscope (Nikon).
  
  Coverslips with surface-attached fruiting bodies were mounted in a custom liquid cell and resubmerged in fresh starvation buffer.
  Prior to all experiments, the liquid cell was placed inside the AFM for a brief period (~\mytilde $\SI{10}{\minute}$) to allow environmental conditions in the unit to stabilize. 
  Experiments on aggregates allowed to develop for less than 24 hours after starvation were performed using Arrow TL1 tipless cantilevers (NanoWorld, $k = 0.02~\text{N/m}$).
  Aggregates allowed to develop for longer than 1 day were measured using tipless FM cantilevers (NanoSensors, TL-FM-20, $k \in [2.8,4.0]~\text{N/m}$) or Cantilever B of BudgetSensors All-in-One tipless probes (BudgetSensors, $k = 2.7~\text{N/m}$). 
  24 hour post-starvation fruiting bodies were measured with Arrow TL1 cantilevers, FM cantilevers from NanoSensors, or Cantilever A ($k = 0.2~\text{N/m}$) of BudgetSensors All-In-One tipless probes.
  For all probes, the manufacturer-specified spring constant, $k$, was used.

  Between all experiments, cantilevers were washed by briefly (\mytilde 30 seconds) submerging in freshly-activated Piranha solution (3:1 v/v mixture $\text{H}_2\text{SO}_4$ and $\text{H}_2\text{O}_2$).
  Note that this treatment both cleaned the cantilever of surface contamination and caused the previously-attached bead to detach from the cantilever.
  Cantilevers were then rinsed with deionized water and allowed to dry before a new bead was attached to their tip (see \ref{supp:methods::cantilever-modification}). 

    \subsubsection{Cantilever Modification}
    \label{supp:methods::cantilever-modification}

    Cantilevers were modified by attaching \mytilde 80 $\mu\text{m}$-diameter solid borosilicate glass beads (Cospheric, Item No. BSGMS-2.2 75-90$\mu\text{m}$-10g) to the end of the cantilever.
    To prevent adhesion between the bead and the fruiting body, prior to attachment, beads were submerged in SigmaCote (Sigma-Aldrich) and placed on a coverslip to air-dry. 
    Meanwhile, a drop of uncured Norland Optical Adhesive 76 (NOA76, Norland Products Inc.) was spread across a coverslip and placed into the AFM in which a freshly-washed cantilever was mounted.
    The surface of the coverslip was then located with the AFM before the cantilever was then manually raised before being moved into contact with the uncured glue.
    After a 1-2 minute incubation period where the glue was allowed to establish contact with the cantilever, the cantilever was raised and the glue-smeared coverslip was replaced with the aforementioned bead coverslip.
    A bead on the coverslip was then aligned with the cantilever tip before the tip was gradually lowered down until bead-cantilever contact was detected.
    The glue was then cured by an approximately 2 minute-long exposure to light from an ultraviolet-light-emitting LED mounted on the AFM stage before being transferred to a 50\degree{}C oven where the glue was allowed to finish curing overnight. 
    
    \subsection{Contact Model}
    \label{supp:methods::contact-model}

  We model the contact between the cantilever-attached-bead and fruiting body as non-adhesive contact between a sphere (the bead) and a hemiellipse (the fruiting body).
  Because of the symmetry brought on by the sphere, such contact can be treated as axisymmetric about the major and minor axes of the hemiellipse and is thus simply the case of Hertzian contact between two spheres with the addition of a small correction factor ($\beta$ in Eqn. \ref{eqn:supp::hertz}) to account for the geometry of the ellipse, \cite{johnson1987contact}.
    \begin{equation}
    F = \frac{4\sqrt{R_c}}{3(1-\nu^2)\beta}E\delta^{3/2},
    \label{eqn:supp::hertz}
  \end{equation}
where $F$ is force, $\nu$ is the Poisson's ratio (assumed $\frac{1}{2}$ for all samples), $\delta$ is the compression depth, $E$ is the Young's modulus, and $R_c$ is the contact radius.
  For details on the determination of $R_c$, see \ref{supp:methods::shape-characterization}, below. 
  Similarly, the solutions for axisymmetric contact between rigid indenters and viscoelastic half-spaces derived by Lee and Radok \cite{lee1960contact} also hold in our geometry. 

  The contact point between the cantilever-attached bead and aggregate of study, and thus the ``zero'' level of sample compression, $\delta$, was found for each experiment using the method described in \cite{chang2014automated}.
  
  \subsubsection{Shape Characterization}
  \label{supp:methods::shape-characterization}

  \begin{figure}[h]
      \includegraphics[width=0.5\textwidth]{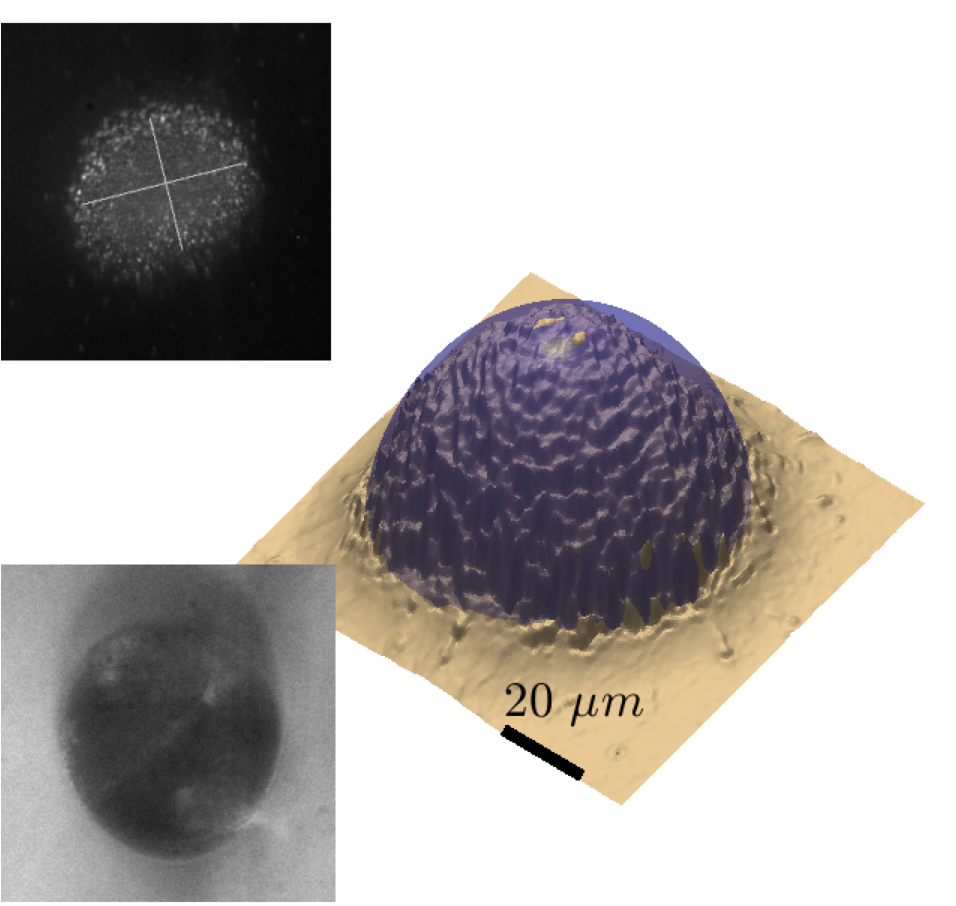}
      \caption{Example microscope images of a fruiting body's base (with major/minor elliptical axes labeled in white, top) and the bead attached to the AFM cantilever, bottom. 
      Main image (tan) is a three-dimensional scan of the same fruiting body, imaged with a Keyence VK-X1000. Purple is elliptical model of the fruiting bodies shape made from the method described in \ref{supp:methods::shape-characterization}.}
      \label{fig:supp::shapechar-examplegeo}
  \end{figure}
  
  Before each experiment, the midline of the bead attached to the cantilever was brought into focus on the AFM's attached microscope and imaged.
  It's radius was then measured manually using Fiji \cite{schindelin2012fiji}.
  
  The shape of each fruiting body was determined by optical measurement of the dimensions at the base followed by measurement of the height by detecting the position of the coverslip surface after each experiment.
  We used the inverted optical microscope to which the AFM was mounted to capture a single image of the fruiting body's base.
  The major and minor diameters were measured from these images manually using Fiji \cite{schindelin2012fiji}.

  The height was measured by using the AFM to detect the coverslip surface immediately next to the fruiting body.
  After each experiment, the microscope stage was moved manually until the first bare coverslip surface was found.
  The AFM cantilever was then lowered gradually until the surface was detected using the same algorithm used to find the top of each fruiting body.
  The height was then taken as the difference between these two positions. 
    
  Given these four values \textemdash{} the major and minor diameters of the fruiting body's base and it's height along with the diameter, we were able to compute the contact radius for each fruiting body/bead combination using the following formulae (see \cite{johnson1987contact}).
  First, we adjust each fruiting body's measured base radius to the radius of curvature along that axis,
  \begin{equation}
    R_\text{maj/min} = \frac{h_{\text{fb}}^2}{R_{\text{maj/min,base}}},
  \end{equation}
  where $h_{\text{fb}}$ is the measured height and $R_{\text{maj/min,base}}$ is the measured major (minor) radius of the fruiting body base.
  We then use these to compute the contact radius ($R_{c}$) along with the two pseudo-radii ($R_a$, $R_b$) that will be used to compute the correction factor, $\beta$,
  \begin{align}
    R_a &= \frac{1}{\alpha - \gamma}, \\
    R_b &= \frac{1}{\alpha + \gamma},  \\
    R_c &= \sqrt{R_{a}R_{b}},
  \end{align}
  where
  \begin{align}
    \alpha &= \frac{1}{2}(\frac{2}{R_{\text{bead}}} + \frac{1}{R_{\text{maj}}} + \frac{1}{R_{\text{min}}}) \\
    \gamma &= \frac{1}{2}\sqrt{(\frac{1}{R_{\text{maj}}} - \frac{1}{R_\text{min}})^2}
  \end{align}
  Finally, we compute the correction factor from $R_a$ and $R_b$,
  \begin{equation}
    \beta = 1 - ((\frac{R_a}{R_b})^{0.0684} - 1)^{1.531}
  \end{equation}

  To validate this approach, we performed the above-described procedure on fruiting bodies and then imaged the surface of these same fruiting bodies on a Keyence Confocal VK-X1000 microscope.
  We then compared the surface predicted by our AFM measurements to that of the VK-X1000.
  As can be seen in Fig. \ref{fig:supp::shapechar-examplegeo}, aside from deviations near the base due to the contact angle of the fruiting body with the coverslip, our approximation matches extremely well.
  This deviation, while it would cause a small (estimated 3-5\%) error in the contact radius, it is also systematic insofar as all fruiting bodies have this same contact angle.
  Thus, while this approximation may cause a minor under-approximation of the absolute value of the modulus, our values are internally-consistent and quantitative. 

  \subsubsection{Stabilization}
  \label{supp:methods::afm-stabilization}

  \begin{figure}[h]
      \centering
      \begin{subfigure}[b]{0.3\textwidth}
        \centering
        \includegraphics{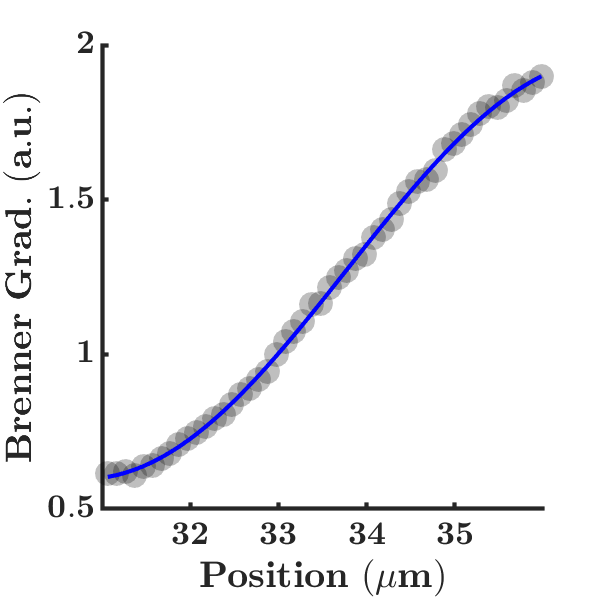}
      \end{subfigure}
      \begin{subfigure}[b]{0.3\textwidth}
        \centering
        \includegraphics{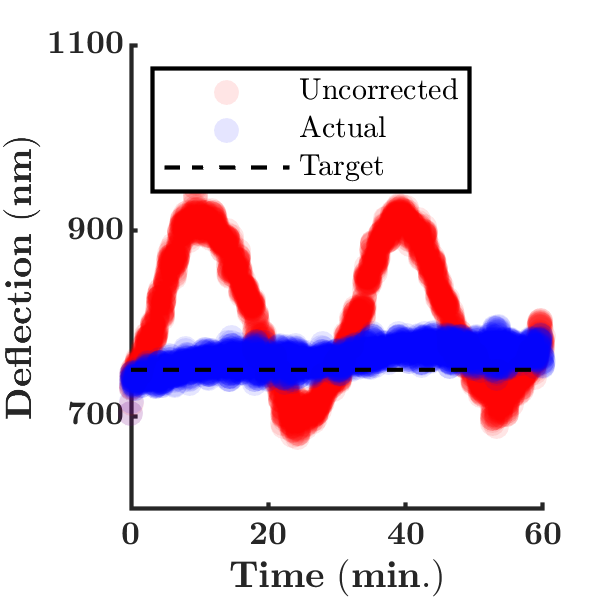}
      \end{subfigure}
      \caption{\textbf{(a)} Example fit of Eqn. \ref{eqn:supp::stable-sinc} (blue) to a the Brenner gradient calculated from an image stack of fixed beads (black) as described in \ref{supp:methods::afm-stabilization}
        \textbf{(b)} Results from an experiment to validate our stabilization technique. A target cantilever deflection of \SI{750}{\nano\meter} was applied at time zero and maintained using only the stabilization method described in \ref{supp:methods::afm-stabilization}. Uncorrected time series was calculated by taking the measured deflection (``Actual'') and adding the motor movements applied by the controller to it.}.
      \label{fig:supp::stabilization}
  \end{figure}
  
  To minimize the effect of mechanical drift on our experiments, we employed an active stabilization mechanism that maintained the coverslip surface in the same reference frame as the AFM tip.
  This stabilization mechanism works by attaching small (\mytilde~$\SI{7}{\micro\meter}$) glass beads to the underside of the sample coverslip, opposite the aggregate to be measured to serve as fiducial markers of the vertical drift in the system.
  Prior to the start of the experiment, a bead directly opposite the aggregate to be measured is brought into focus in the microscope.
  The microscope objective is then swept up and down relative to this position and at each point in the stack, the Brenner gradient \cite{brenner1976automated} of the resulting image is calculated.
  The resulting Brenner gradient vs. position graph is then fit to Eqn. \ref{eqn:supp::stable-sinc} and inverted to yield an empirical relationship between the focus of the reference bead and vertical position of the coverslip relative to the objective,
  \begin{equation}
    \text{Brenner}(z) = A_{1}\sinc(z-A_{2})\sin(A_{3}(z-A_{2})) + A_{4}.
    \label{eqn:supp::stable-sinc}
  \end{equation}

  During an experiment, the bead is then maintained in focus by a PI controller that modulates the microscope objective up and down to match the (implied) vertical drift of the AFM.
  These movements are then applied in tandem to the motor controlling the position of the AFM cantilever, thus maintaining the coverslip surface and cantilever tip position in the same reference frame throughout an experiment.
  To validate this approach, we used the AFM to apply a small force (fixed cantilever deflection) to a bare coverslip and used the active stabilization to maintain that force over time.
  Example results are shown in Fig. \ref{fig:supp::stabilization}(b).
  
  \subsection{Creep Experiments}
  \label{supp:methods::creep-expts}
  
  Creep experiments were performed by indenting each fruiting body to a pre-selected force and held for 10 minutes.
  The force was maintained by a PI controller, running at closed loop frequency 20 Hz and sampling data at 1 kHz, that was tuned for each age by manually adjusting parameters on trial fruiting bodies until parameters yielding the (apparently) fastest rise time without overshoot were found.
  Because of the vast differences in stiffness between fruiting bodies at different times post-starvation, the set force was chosen such that the level of compression would be within the linear viscoelastic regime for each fruiting body (approximately 1-5 $\mu{}\text{m}$).
  The set force for each age is listed in Table \ref{tbl:supp::creepSetForces}.

  \begin{table}[h]
    \begin{tabular}{ |c|c| } \hline
      \textbf{Hours Post-Starvation} & \textbf{Force (nN)} \\
      \hline
      12 & 60 \\
      15 & 60 \\
      18 & 60 \\
      21 & 60 \\
      24 & 350 \\
      48 & 350 \\
      72 & 350 \\
      96 & 350 \\
      168 & 350 \\
      \hline
    \end{tabular}
    \caption{Set forces used in creep experiments for each experimental timepoint.}
    \label{tbl:supp::creepSetForces}
  \end{table}

  \subsubsection{Model Fitting}
  \label{supp:methods::creep-model-fitting}

  To extract viscoelastic model parameters from creep experiments, we modified the algorithm described in \cite{efremov2017measuring} to the Lee-Radok solution for viscoelastic sample compression where sample-probe contact area only increases over the course of the experiment,
  \begin{equation}
    \delta^{3/2}(F,t) = \frac{9\beta}{16\sqrt{R_c}}\int_0^t J(t-\tau)\frac{dF}{d\tau}d\tau.
    \label{eqn:supp::leeRadok}
  \end{equation}
  Note that we have slightly modified Lee and Radok's original formulation by introducing the shape correction factor, $\beta$, described earlier (see \ref{supp:methods::contact-model}).

  Compression ($\delta$ in Eqn. \ref{eqn:supp::leeRadok}) vs. time traces for each experiment were generated by fitting the contact point using the method of \cite{chang2014automated} and subtracting the instantaneous cantilever position at each recorded time from this value.
  Force ($F$ in Eqn. \ref{eqn:supp::leeRadok}) vs. time traces were generated by smoothing the QPD values recorded by the controller sampling loop with a second order Savitzky-Golay filter over a 1 second time window.
  This same filter was then used to esimate the time-derivative of the force used in Eqn. \ref{eqn:supp::leeRadok}.
  Data was then averaged into 100 log-spaced bins between the experimental sampling time ($\SI{0.05}{\second}$) and total time ($\SI{600}{\second}$). 
  
  Model parameters for $J(t)$ were extracted from these compression and force traces by numerically evaluating Eqn. \ref{eqn:supp::leeRadok} and minimizing the least-squares error between the predicted and experimental indentation time series.
  Minimization was done using the trust region reflective algorithm of MATLAB's \texttt{lsqcurvefit} function.
  All computations were done in MATLAB R2019a (Mathworks, Inc.). 
  
  \subsection{Frequency Domain Experiments}
  \label{supp:methods::oscillatory-expts}

  Measurements of the dynamic modulus were conducted by the methods of \cite{mahaffy2004quantitative,huang2004measurements}.
  For droplets $\ge\SI{24}{\hour}$ post-starvation, we performed indentation-controlled frequency sweeps about a  compression of $\delta_{0} = \SI{1}{\micro\meter}$.
  For each experiment, we compressed the sample to $\delta_{0}$ for 10 minutes prior to beginning oscillations.
  Oscillations, $\widetilde{\delta}=\SI{100}{\nano\meter}$ were then imposed in 20 log-spaced steps between $\SI{1}{\milli\hertz}$ and $\SI{10}{\hertz}$.
  For each frequency step, two cycles were imposed and the resulting compression and force time series recorded at a rate of 100 samples/cycle.
  
  For droplets 12 hours post-starvation, we found that the high degree with which these droplets relaxed stresses prevented us from making reliable measurements. 
  We thus used force-controlled frequency sweeps about a fixed force, $F_{0}=\SI{60}{\nano\newton}$, with oscillating force $\widetilde{F}=\SI{2}{\nano\newton}$.
  Sweeps were again imposed in 20 log-spaced steps between $\SI{1}{\milli\hertz}$ and $\SI{10}{\hertz}$ and all data recorded at a rate of 100 samples/cycle.

\begin{figure}
    \includegraphics[width=\textwidth]{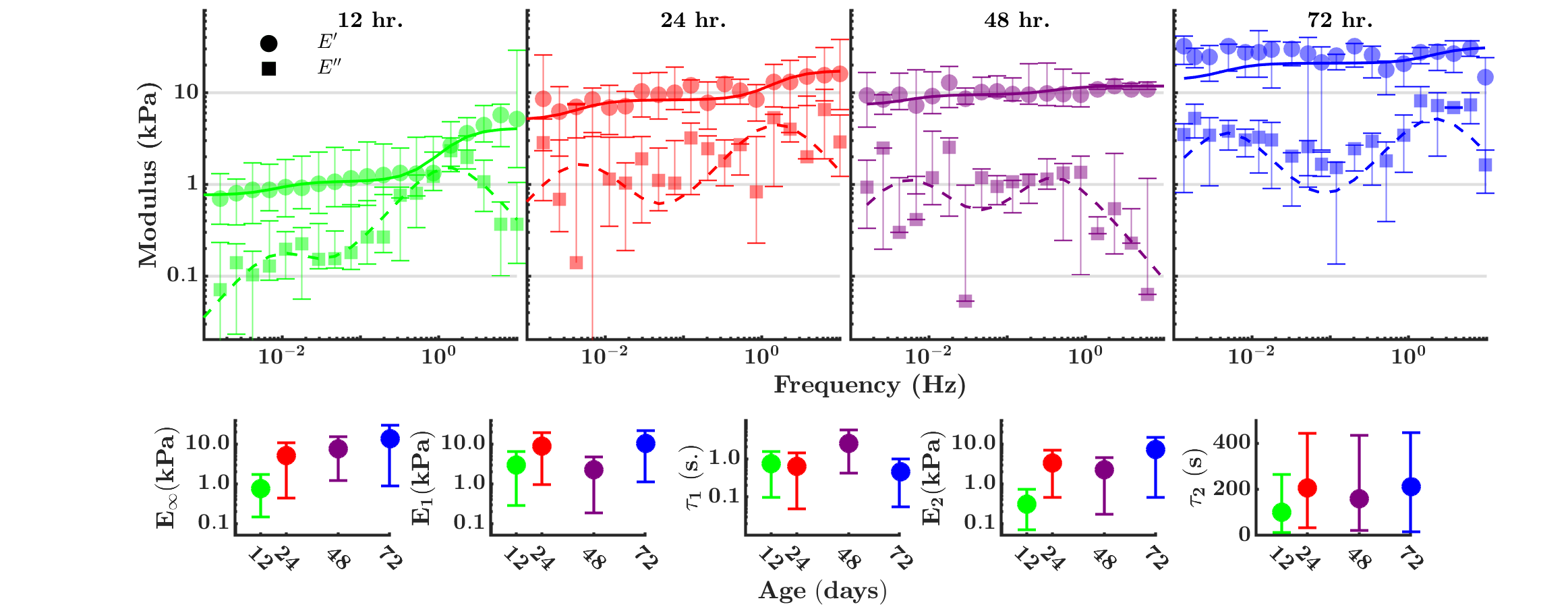}
    \caption{Dynamic modulus measurements, as shown in Fig. \ref{fig:main::freqSweep} but only considering a one-term Taylor expansion instead of the two-term expansion used throughout this work.}
    \label{fig:supp::oneordertaylor}
  \end{figure}

  For each frequency step, the two cycles were fit to a 1-term fourier series.
  The parameters from that fit were then extracted to yield values for the exact force and indentation offsets ($F_{0},~\delta_{0}$, respectively) and amplitudes, $\widetilde{F},~\widetilde{\delta}$.
  We then follow a similar approach to both \cite{mahaffy2004quantitative,huang2004measurements} and consider the Hertz model for compressions and forces of the form $\delta = \delta_{0}+\widetilde{\delta}e^{i\omega{}t}, F = F_{0}+\widetilde{F}e^{i\omega{}t}$ and consider Taylor expansions of Eqn. \ref{eqn:main::hertz} in the compression.
  Considering terms up to second order yields
  \begin{align}
    F_{0} &= \frac{16\sqrt{R_c}}{9\beta}E\delta_{0}^{3/2} \label{eqn:supp::taylorexpansionf0}, \\
    \widetilde{F}(\omega) &= \frac{16\sqrt{R_c}}{9\beta}E^{*}(2\sqrt{\delta_0}\widetilde{\delta}+\widetilde{\delta}^{2}/\sqrt{\delta_{0}}). \label{eqn:supp::taylorexpansionfosc} \\
  \end{align}
  Rearranging Eqn. \ref{eqn:supp::taylorexpansionfosc} for $E^{*}$ then yields:
  \begin{equation}
    E^{*}(\omega) = \frac{9\beta\widetilde{F}(\omega)}{16\sqrt{R_c}(2\sqrt{\delta_0}\widetilde{\delta}+\widetilde{\delta}^{2}/\sqrt{\delta_{0}})}
    \label{eqn:supp::taylorexpansionfull}
  \end{equation}
  Because of the relative scale of our offset and amplitude, we consider terms up to second order and report values using the formulas given above.
  Considering only the first term does not affect our results, as we show in Fig. \ref{fig:supp::oneordertaylor}

\section{Viscoelastic Modeling}
\label{supp:viscoelastic-modeling}

We model the apparent viscoelasticity of \textit{M. xanthus} fruiting bodies using a 2-element Prony series. 
This model came about empirically as the simplest model that could describe the data both individually (as in Fig. \ref{fig:main::exampleExpts}) and in aggregate (as in Fig. \ref{fig:main::freqSweep}) where we observe 2 relaxation timescales and an apparent equilibrium modulus at low frequency.
Similarly, in creep we observe two retardation timescales and an apparent glass compliance, characteristic of a 2-element Generalized Kelvin-Voigt model in series with a lone spring. 
Like other Prony series representations of viscoelastic behavior, the conjugacy of the two aforementioned models allows us to interconvert between their respective relaxation moduli and creep compliances.
However unlike more complex (\textit{e.g.} $N > 2$) models whose interconversion can often only be done numerically \cite{park1999methods}, we can convert exactly between these two models using the ``method of residues'' described by Tschoegel \cite{tschoegl2012phenomenological}.

Starting with the Laplace-domain relaxance of a 2-element generalized Maxwell model, we seek the retardation times and creep compliances of its conjugate Kelvin-Voigt model,
\begin{equation}
  \label{eqn:supp::gmm_relaxance}
  Q_{\text{MW}}(s) = G_{\text{eq}} + \frac{G_{1}\tau_{1}s}{1 + \tau_{1}s} + \frac{G_{2}\tau_{2}s}{1 + \tau_{2}s}.
\end{equation}
The dynamic modulus can be arrived at by simply replacing the Laplace domain variable $s$ with complex frequency $i\omega$,
\begin{equation}
  \label{eqn:supp::gmm_dynmod}
  G^{*} = G_{\text{eq}} + \frac{G_{1}\tau_{1}i\omega}{1 + \tau_{1}i\omega} + \frac{G_{2}\tau_{2}i\omega}{1 + \tau_{2}i\omega}.
\end{equation}
The retardance by inversion of relaxance,
\begin{equation}
  \label{eqn:supp::gmm_retardance}
  U_{\text{MW}}(s) = \frac{1}{Q_{\text{MW}}(s)} = \frac{\tau_{1}\tau_{2}s^{2} + (\tau_{1}+\tau_{2})s + 1}{s^{2}(G_1\tau_1\tau_2 + G_2\tau_1\tau_2 + G_{\text{eq}}\tau_1\tau_2) + s(G_1\tau_1 + G_2\tau_2+G_{\text{eq}}\tau_1+G_{\text{eq}}\tau_2) + G_{\text{eq}}}.
\end{equation}

From this we can compute the retardation times ($\lambda_n$) of the conjugate Kelvin-Voigt model from the poles of Eqn. \ref{eqn:supp::gmm_retardance} and the recognition that $s_n = -1/\lambda_n$ \cite{tschoegl2012phenomenological}.
These are
\begin{equation}
  \label{eqn:supp::gmm_retardationtimes}
  \lambda_n = \frac{2(G_1\tau_1\tau_2 + G_2\tau_1\tau_2 + G_{\text{eq}}\tau_1\tau_2)}{G_1\tau_1 + G_2\tau_2 + G_{\text{eq}}\tau_1 + G_{\text{eq}}\tau_2 \pm \sqrt{\alpha}},
\end{equation}
where
\begin{equation}
  \label{eqn:supp::gmm_retardation_alpha}
  \alpha = (G_1\tau_1)^2+2G_1G_2\tau_1\tau_2 + 2G_1G_{\text{eq}}\tau_1^2 - 2G_1G_{\text{eq}}\tau_1\tau_2 + (G_2\tau_2)^2 - 2G_2G_{\text{eq}}\tau_1\tau_2 + 2G_{2}G_{\text{eq}}\tau_2^2 + (G_{\text{eq}}\tau_1)^2 - 2G_{\text{eq}}^2\tau_1\tau_2 + (G_{\text{eq}}\tau_2)^2.
\end{equation}
Similarly, the compliances can be computed by the relation $J_n = \tau_{n}\text{Res}_n(U)$ where $\text{Res}_{n}$ is the residue associated with the $n^{\text{th}}$ pole \cite{tschoegl2012phenomenological}.

Applying Eqn. 4.5-31 of \cite{tschoegl2012phenomenological} to compute the residues and the aforementioned relation yields
\begin{align}
  \label{eqn:supp::gmm_compliance1}
  J_{1} &= \frac{(G_1\tau_1 + 2G_2\tau_1 - G_2\tau_2 + G_{\text{eq}}\tau_1 - G_{\text{eq}}\tau_2 + \sqrt{\alpha}) (2G_1\tau_2 - G_1\tau_1+G_2\tau_2-G_{\text{eq}}\tau_1+G_{\text{eq}}\tau_2 + \sqrt{\alpha})}{2\sqrt{\alpha}(G_1+G_2+G_{\text{eq}})(G_1\tau_1 + G_2\tau_2 + G_{\text{eq}}\tau_1 + G_{\text{eq}}\tau_2 - \sqrt{\alpha})}, \\
  \label{eqn:supp::gmm_compliance2}
  J_{2} &= \frac{(G_1\tau_1 + 2G_1\tau_2 - G_2\tau_2 + G_{\text{eq}}\tau_1 - G_{\text{eq}}\tau_2 + \sqrt{\alpha}) (G_1\tau_1 + 2G_2\tau_1 - G_2\tau_2 + G_{\text{eq}}\tau_1 - G_{\text{eq}}\tau_1 - \sqrt{\alpha})}{2\sqrt{\alpha}(G_1+G_2+G_{\text{eq}})(G_1\tau_1 + G_2\tau_2 + G_{\text{eq}}\tau_1 + G_{\text{eq}}\tau_2 + \sqrt{\alpha})},
\end{align}
where $\alpha$ is the same as that in Eqn. \ref{eqn:supp::gmm_retardation_alpha}. 
The glass compliance, $J_0$ can be found from the high-frequency limit of Eqn. \ref{eqn:supp::gmm_retardance},
\begin{equation}
  \label{eqn:supp::gmm_glasscompliance}
  J_0 = \lim_{s \to \infty} U_{\text{MW}}(s) = \frac{1}{G_{\text{eq}} + G_1 + G_2}.
\end{equation}

For the reverse conversion of creep compliances/retardation times to moduli/relaxation times, we begin with the retardance of the 2-element generalized Kelvin-Voigt model and reapply the above procedure to find the relaxation times from the poles of the relaxance,
\begin{align}
  \label{eqn:supp::gkv_retardance}
  U_{\text{KV}}(s) &= J_0 + \frac{J_1}{\lambda_{1}s + 1} + \frac{J_1}{\lambda_{2}s + 1}, \\
  \label{eqn:supp::gkv_relaxance}
  Q_{\text{KV}}(s) &= \frac{1}{U_{\text{KV}}(s)} = \frac{s^2(\lambda_1\lambda_2) + s(\lambda_1+\lambda_2) + 1}{s^2(J_0\lambda_1\lambda_2) + s(J_1\lambda_2 + J_2\lambda_1 + J_0\lambda_1 + J_0\lambda_2) + (J_1 + J_2 + J_0)}, \\
  \tau_n &= \frac{2J_0\lambda_1\lambda_2}{J_1\lambda_2 + J_2\lambda_1 + J_0\lambda_1 + J_0\lambda_2 \pm \sqrt{\beta}}, \label{eqn:supp::gkv_relax_times}
\end{align}
where
\begin{equation}
  \label{eqn:supp::gkv_beta}
  \beta = (J_1\lambda_1)^2 + 2J_1J_2\lambda_1\lambda_2 - 2J_1J_0\lambda_1\lambda_2 + 2J_1J_0\lambda_2^2 + (J_2\lambda_1)^2 + 2J_2J_0\lambda_1^2 - 2J_2J_0\lambda_1\lambda_2 + (J_0\lambda_1)^2 - 2J_0^2\lambda_1\lambda_2 + (J_0\lambda_2)^2.
\end{equation}
The moduli are found using $G_n = -\tau_n\text{Res}_n(Q)$,
\begin{equation}
  \label{eqn:supp::gkv_moduli}
  G_n = \pm \frac{2J_0\lambda_1\lambda_2(\frac{J_1\lambda_2+J_2\lambda_1+J_0\lambda_1+J_0\lambda_2 \pm \sqrt{\beta}}{2J_0\lambda_1}-1)(\frac{J_1\lambda_2+J_2\lambda_1+J_0\lambda_1+J_0\lambda_2 \pm \sqrt{\beta}}{2J_0\lambda_2}-1)}{\sqrt{\beta}(J_1\lambda_2+J_2\lambda_1+J_0\lambda_1+J_0\lambda_2 \pm \sqrt{\beta})},
\end{equation}
where $\beta$ is given by Eqn. \ref{eqn:supp::gkv_beta}. 

Finally, the equilibrium modulus is the low-frequency limit of the relaxance,
\begin{equation}
  \label{eqn:supp::gkv_eqmod}
  G_{\text{eq}} = \lim_{s \to 0} Q_{\text{KV}}(s) = \frac{1}{J_1 + J_2 + J_0}.
\end{equation}
Note that we can also use Eqn. \ref{eqn:supp::gkv_retardance} to compute the time-dependent creep compliance for this model,
\begin{equation}
  \label{eqn:supp::gkv_creepcompliance1}
  J(t) = \mathscr{L}^{-1}\{\frac{1}{s}U_{\text{KV}}\} = J_0 + J_1(1 - e^{-t/\lambda_1}) + J_2(1 - e^{-t/\lambda_2}),
\end{equation}
or equivalently, defining $q_n = J_n/J_0$,
\begin{equation}
  J(t) = J_0 [ 1 + q_1(1 - e^{-t/\lambda_1}) + q_2(1 - e^{-t/\lambda_2}) ].
\end{equation}

\putbib[Black_MyxoAFM2021_refs]
\end{bibunit}
\end{document}